\documentclass{PoS}

\title{Results for the mass difference between the long- and short- lived K mesons for physical quark masses}

\ShortTitle{$\Delta m_K$ for physical quark masses}

\author{\speaker{Bigeng Wang}\thanks{This work was partially supported by US DOE grant
\#DE-SC0011941 and
used computer time provided by the Innovative and Novel Computational Impact
on Theory and Experiment (INCITE) program. This research used resources of
the Argonne Leadership Computing Facility, which is a DOE Office of Science
User Facility supported under Contract DE-AC02-06CH11357.}\\
        Department of Physics, Columbia University, New York, NY 10027, USA\\
        E-mail: \email{bw2482@columbia.edu}}

\abstract{The two neutral kaon states in nature, the $K_L$ (long-lived) and $K_S$ (short-lived) mesons, are the two time-evolution eigenstates of the $K^0 - \overline{K^0}$ mixing system. The prediction of their mass difference $\Delta m_K$ based on the Standard Model is an important goal of lattice QCD. In this article, I will present preliminary results from a calculation of $\Delta m_K$ performed on an ensemble of $64^3 \times 128$ gauge configurations with inverse lattice spacing of 2.36 GeV and physical quark masses. These new results come from 2.5 times the Monte Carlo statistics used for the result presented in last year's conference. Further discussion of the methods employed and the resulting systematic errors will be given.}

\FullConference{The 36th Annual International Symposium on Lattice Field Theory - LATTICE2018\\
		22-28 July, 2018\\
		Michigan State University, East Lansing, Michigan, USA.}

\begin{document}

\section{Introduction}
The mass difference between $K_L$ and $K_S$ is generated by K meson mixing through $\Delta S = 2$ weak interaction. With its experimental value of $3.483(6) \times 10^{-12}$ MeV measured with sub percentage error, a discrepancy between the prediction based on the Standard Model and this value implies the existence of physics beyond the Standard Model. This quantity is highly non-perturbative and can be calculated using Lattice QCD from first principles. Since 2013 an exploratory calculation on a $16^3 \times 32$ calculation, with unphysical masses ($m_\pi = 421$ MeV) including only connected diagram\cite{jianglei}, the RBC-UKQCD collaborations have been improving the calculation by including disconnected diagrams and extending measurements to finer lattice spacing \cite{jianglei_2}. Our most recent calculation on a $64^3 \times 128$ lattice with physical masses on 59 configurations gives a preliminary result of $\Delta m _k=(5.5 \pm 1.7) \times 10^{-12}$ MeV \cite{Bai:2018mdv}. In this article, an update of the methods and results extending our calculation from 59 to 152 configurations is presented.

\section{Integrated Correlator and $\Delta m_K$}
The $K_L-K_S$ mass difference is expressed as:
\begin{equation}
    \label{eqn:dmk}
    \Delta M_K = 2 Re M_{\overline{0}0} = 2 \mathcal{P} \sum_{n} \frac{\langle \bar{K}^0|H_W|n\rangle\langle n|H_W|K^0\rangle}{m_K-E_n}.
\end{equation}
To evaluate $\Delta m_K$ on an Euclidean space lattice, we evaluate the integrated correlators:\\
\begin{equation}
    \mathcal{A}(T)= \frac{1}{2} \sum^{t_b}_{t_2=t_a} \sum_{t_1=t_a}^{t_b}
    \langle 0 |T \{ \bar{K}^0(t_f)H_W(t_2)H_W(t_1)K^0(t_i) \} | 0 \rangle,
    \label{eqn:intgrate}
\end{equation}
where $H_W$ is the $\Delta S =1 $ effective Hamiltonian:
\begin{equation}
\label{eqn:hamiltonian}
    H_W = \frac{G_F}{\sqrt{2}} \sum_{q,q'=u,c} V_{qd}V^*_{q's}
    (C_1Q_1^{qq'}+C_2Q_2^{qq'}).
\end{equation}
Here the ${Q_i^{qq'}}_{i=1,2}$ are current-current operators, defined as:
\begin{equation}
\label{eqn:q1q2}
Q_1^{qq'} = (\bar{s}_i \gamma^{\mu} (1- \gamma_5)d_i)(\bar{q}_j \gamma^{\mu}(1-\gamma_5)q_j'), \quad
Q_2^{qq'} = (\bar{s}_i \gamma^{\mu} (1- \gamma_5)d_j)(\bar{q}_j \gamma^{\mu}(1-\gamma_5)q_i'),
\end{equation}
and $V_{q_a q_b}$ are the usual CKM matrix elements and $C_i$ are Wilson coefficients.

If we insert a complete set of intermediate states, we identify the coefficient of the term linear in the size of integration box $T = t_b - t_a + 1$ as proportional to the expression for $\Delta m_K$ given in Equation \ref{eqn:dmk}:
\begin{equation}
    \mathcal{A}(T)=N_K^2 e^{-m_K(t_f-t_i)} \sum_{n} \frac{\langle \bar{K}^0|H_W|n\rangle\langle n|H_W|K^0\rangle}{m_K-E_n} \{-T+\frac{e^{(m_K-E_n)T}-1}{m_K-E_n} \}.
\end{equation}
\\
Before doing a linear fitting with respect to $T$, the second term in the curly bracket has to be removed. For an intermediate state $| n \rangle$ with energy $E_n$ larger than $m_K$, for large enough $T$, the contribution from the second term is negligible. For a state $| n \rangle$ with energy $E_n$ smaller than or close to $m_K$, we need to subtract its contribution. 

In our case of physical quark masses, $|0\rangle$, $|\pi\pi\rangle$, $|\eta\rangle$ and  $|\pi\rangle$ states need to be subtracted. With the freedom of adding the operators $\overline{s}d$ and $\overline{s} \gamma_5 d$ to the weak Hamiltonian with properly chosen coefficients $c_s$ and $c_p$, we are able to remove two of the contributions. Here we choose $c_s$ and $c_p$ to satisfy Equation \ref{eqn:cpcs} so that contributions from $|0\rangle$ and $|\eta\rangle$ will vanish: 
\begin{equation}
\label{eqn:cpcs}
    \langle 0 | H_W - c_p \bar{s}\gamma_5 d| K^0\rangle=0, \quad
    \langle \eta | H_W - c_s \bar{s}d| K^0\rangle=0.
\end{equation}
As a result, the original $\Delta S = 1$ effective weak Hamiltonian in Equation \ref{eqn:hamiltonian} and the current-current operators should be modified to be :
\begin{equation}
   Q_i'= Q_i - c_{pi} \bar{s}\gamma_5 d - c_{si} \bar{s}d
\end{equation}
with $c_{pi}$ and $c_{si}$ are calculated on lattice using Equation \ref{eqn:cscp_calc}.\\
\begin{equation}
\label{eqn:cscp_calc}
    c_{si}=\frac{\langle \eta | Q_i | K^0\rangle}{\langle \eta | \overline{s}d | K^0\rangle},  \quad
    c_{pi}=\frac{\langle 0 | Q_i | K^0\rangle}{\langle 0| \overline{s}\gamma_5 d | K^0\rangle}.
\end{equation}

\begin{figure}
    \centering
    \includegraphics[width=0.7\textwidth]{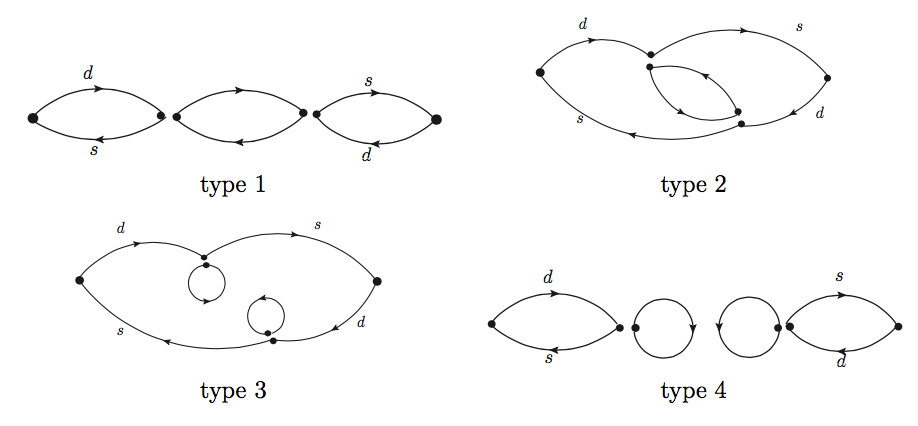}
    \caption{Four types of contractions in the 4-point correlators with $Q_1$ and $Q_2$.}
    \label{fig:contract_q}
\end{figure}

For contractions among $Q_i$, there are four types of diagrams to be evaluated, as shown in Figure \ref{fig:contract_q}. In addition, there are "mixed" diagrams from the contractions between the $\bar{s}d$, $\bar{s}\gamma_5d$ and $Q_i$ operators, having similar topologies to type 3 and type 4 contractions. 

The GIM mechanism removes both quadratic and logarithmic divergences that might otherwise be expected as the two $H_W$ operators approach each other. We therefore include the charm quark in our calculation and as a result always have the difference between up and charm quark propagators for every charge $+2/3$ quark line.

\begin{figure}
    \centering
    \includegraphics[width=0.25\textwidth]{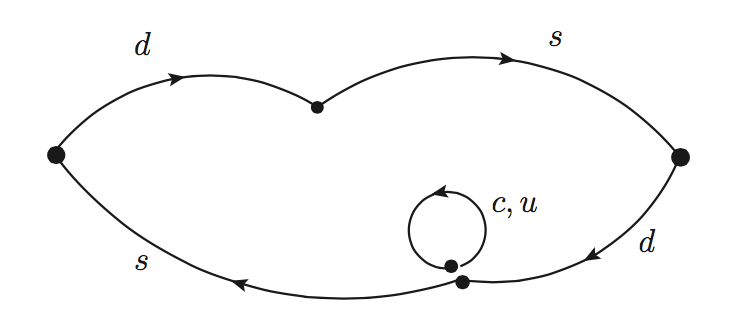}
    \includegraphics[width=0.3\textwidth]{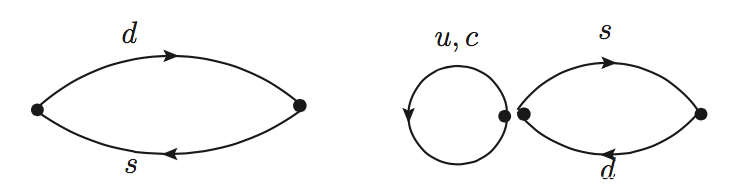}\\
    \includegraphics[width=0.25\textwidth]{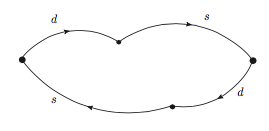}
    \includegraphics[width=0.3\textwidth]{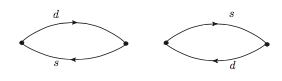}
    \caption{The "mixed" contractions in the 4-point correlators. The top two are products of the $c_{si}\bar{s}d$ and $c_{pi} \bar{s}\gamma_5d$ with the $Q_i$ and the bottom two are products of the $c_{si}\bar{s}d$ and $c_{pi} \bar{s}\gamma_5d$. The diagrams on the left are similar to type-3 diagrams and the diagrams on the right are similar to type-4 diagrams. }
    \label{fig:contract_sd}
\end{figure}

\section{From lattice results to physical $\Delta m_K$}

The fitting of the integrated correlator in Equation \ref{eqn:intgrate} further breaks into fitting of the integrated correlator with $Q_1$ and $Q_2$:

\begin{equation}
    \mathcal{A}_{ij}(T)=N_K^2 e^{-m_K(t_f-t_i)} \sum_{n} \frac{\langle \bar{K}^0|Q_i|n\rangle\langle n|Q_j|K^0\rangle}{m_K-E_n} \{-T+\frac{e^{(m_K-E_n)T}-1}{m_K-E_n} \}.
    \label{eqn:integrate_ij}
\end{equation}

Considering the GIM mechanism, the relationship between $\mathcal{A}_{ij}(T)$ in Equation \ref{eqn:integrate_ij} and $\mathcal{A}(T)$ in Equation \ref{eqn:intgrate} is:
\begin{equation}
    \mathcal{A}(T) = \lambda_u^2 \sum_{i,j=1,2} C_i C_j \mathcal{A}_{ij}(T),
\end{equation}
where the $C_i$ are Wilson coefficients and $\lambda_u = V_{ud}V_{us}^{*}$.

We fit each $\mathcal{A}_{ij}(T)$ separately and obtain the $k_{ij}$, coefficient of the linear term of $T$. The value of $\Delta m_K$ from the lattice should be:

\begin{equation}
    \Delta m_K^{lat} = \frac{G_F^2}{2}\lambda_u^2 \sum_{i,j=1,2} (-2)\times C_i^{lat} C_j^{lat} k_{ij}.   
\end{equation}

We obtain the Wilson coefficients of the operators $C_i^{lat}$ in three steps \cite{NPR} \cite{NPR2}:
\begin{itemize}
        \item 
        Non-perturbative Renormalization: from lattice to RI-SMOM
        \item
        Perturbation theory: from RI-SMOM to $\overline{MS}$
        \item
        Perturbation theory: Wilson coefficients in the $\overline{MS}$ scheme
        
    \end{itemize}
    So $C_1^{lat}$ and $C_2^{lat}$ can be expressed as: 
    \begin{equation}
        C_i^{lat}= C_a^{\overline{MS}} (1 + \Delta r )^{RI \rightarrow \overline{MS}}_{ab} Z^{lat \rightarrow RI}_{bi}.
    \end{equation}

\section{Sample AMA and super-jackknife method}
We use sample All Mode Averaging(AMA) method to reduce the computational cost\cite{Sample AMA}. The usual AMA correction is applied on each configuration, among different time slices. In contrast the sample AMA correction is applied among configurations: on most configurations, quantities are calculated with a CG stopping residual of $10^{-4}$("sloppy"). On the other configurations the same quantities are calculated with a CG stopping residual of both  $10^{-4}$("sloppy") and $10^{-8}$("exact"). The differences between "sloppy" and "exact" measurements are used as corrections to the "sloppy" only configurations.

In our case, we have data for type-3 and type-4 diagrams, three-point and two-point functions from both "sloppy" measurements and corrections. We firstly jackknife the "sloppy" and correction data separately and then use the super-jackknife method to estimate the error of the super-jackknife samples. For a certain quantity $Y$, a pion correlator as an example, from the $N_s$ "sloppy" measurements $\big\{y_i \big\}_{i = 1, ... , N_s}$, we obtain the jackknife "sloppy" ensemble $\big\{Y_i\big\}_{i = 1, ... , N_s}$ with $Y_i = \frac{1}{N_s-1} \sum_{j \ne i} y_j$. Similarly, from the $N_c$ corrections $\big\{\Delta y_i \big\}_{i = 1, ... ,N_c}$, we obtain the jackknife correction ensemble $\big\{\Delta Y_i\big\}_{i = 1, ... , N_c}$ with $\Delta Y_i = \frac{1}{N_c-1} \sum_{j \ne i} \Delta y_j$. We then combine the two jackknife ensembles to form a super-jackknife ensemble $\big\{ Y_k' \big\}_{k = 1, ... , N_s + N_c}$ with $N_s + N_c$ elements, where: 
    \begin{equation}
        Y_k' = Y_k + \overline{\Delta Y}, \quad k = 1, ... N_s  
    \end{equation}
    \begin{equation}
        Y_k' = \overline{Y} + \Delta Y_{k-N_s}, \quad k = N_s + 1, ... N_s + N_c 
    \end{equation}
where $\overline{\Delta Y} = \frac{1}{N_c} \sum_{j=1}^{N_c} \Delta Y_j$ is the mean value of the corrections and $\overline{Y} = \frac{1}{N_s} \sum_{i=1}^{N_s} Y_i$ is the mean value of the "sloppy" measurements.
\section{Lattice calculation and results}
The calculation was performed on a $64^3 \times 128 \times 12$ lattice with 2+1 flavors of M$\ddot{o}$bius DWF and the Iwasaki gauge action with physical pion mass (136 MeV) and inverse lattice spacing $a^{-1} = 2.36$ GeV. The inputs parameters are listed in Table \ref{tab:inputs}. We will compare results presented in Lattice 2017 \cite{Bai:2018mdv} with our updated results. We have in total 152 configurations, among which 116 configurations are "sloppy" and 36 configurations are used for corrections.  In the tables below, we refer to the updated data set as "new 152" and data set used in 2017 as "old 59".
\begin{table}[]
    \centering
    \begin{tabular}{ | c  | c | c | c | c |}
    \hline
   $\beta$ & $am_l$  & $am_h$ & $\alpha=b+c$ & $L_s$\\ \hline
 2.25 & 0.0006203 & 0.02539 & 2.0 &  12 \\ \hline
	
    \end{tabular}
    \caption{Input parameters of the lattice calculation.}
    \label{tab:inputs}
\end{table}

The masses from fitting two-point correlators are included in Table \ref{tab:2pt}. Amplitudes and coefficients for subtractions are listed in Table \ref{tab:3pt}, Table \ref{tab:sub}, and Table \ref{tab:ktopipi}. These results are consistent within errors. As the statistics increase, the errors scale approximately as $\frac{1}{\sqrt{N}}$, where $N$ is the number of total measurements.  \\

\begin{table}[]
    \centering
    \begin{tabular}{c|c|c|c|c } \hline
      Data Set   &$K^0$& $\pi$ &  $\eta$ & $\pi\pi_{I=0}$\\ \hline
      new 152 &0.2104(1) & 0.0574(1) & 0.258(2)& 0.1138(5) \\ \hline
      old 59 &0.2105(2) & 0.0576(1) & 0.290(29)& 0.1137(8) \\ \hline
    \end{tabular}
    \caption{Fitting results for meson masses and $\pi-\pi$ energy in lattice units ($a^{-1} = 2.36 $ GeV)}
    \label{tab:2pt}
\end{table}

\begin{table}[!htb]
    \centering
    \begin{tabular}{c|c|c|c|c} \hline
       Data Set &$\langle \pi | Q_1 | K^0 \rangle $ & $\langle \pi | Q_2 | K^0 \rangle $ & $\langle 0 | Q_1 | K^0 \rangle $ & $\langle 0 | Q_2 | K^0 \rangle $\\ \hline
       new 152 &$-5.02(3) \times 10^{-4}$ & $1.407(4) \times 10^{-3}$ &$-1.284(3) \times 10^{-2}$ & $2.449(4) \times 10^{-2}$ \\ \hline
        old 59 &$-5.08(5) \times 10^{-4}$ & $1.407(8) \times 10^{-3}$ &$-1.289(4) \times 10^{-2}$ & $2.454(7) \times 10^{-2}$ \\ \hline
    \end{tabular}
    \caption{ The $K^0$ to $\pi$ matrix element and the $K^0$ to vacuum matrix element, without subtracting the $\bar{s}d$ operator.}
    \label{tab:3pt}
\end{table}

\begin{table}[!htb]
    \centering
    \begin{tabular}{c|c|c|c|c} \hline
       Data Set  &$c_{s1}$& $c_{s2}$ & $c_{p1}$ & $c_{p2}$  \\ \hline
        new 152& $2.13(33) \times 10^{-4}$ & $-3.16(25) \times 10^{-4} $& $-1.472(2) \times 10^{-4}$ & $2.807(2) \times 10^{-4}$  \\ \hline
        old 59& $1.53(64) \times 10^{-4}$ & $-2.77(42) \times 10^{-4} $& $-1.476(3) \times 10^{-4}$ & $2.811(3) \times 10^{-4}$  \\ \hline
    \end{tabular}
    \caption{The subtraction coefficients for the scalar and pseudo-scalar operator.}
    \label{tab:sub}
\end{table}

\begin{table}[!htb]
    \centering
    \begin{tabular}{c|c|c|c|c} \hline
       Data Set &$\langle \pi\pi_{I=2} | Q_1 | K^0 \rangle $ & $\langle \pi\pi_{I=2} | Q_2 | K^0 \rangle $ & $\langle \pi\pi_{I=0}| Q_1 | K^0 \rangle $ & $\langle \pi\pi_{I=0} | Q_2 | K^0 \rangle $\\ \hline
        new 152 &$1.473(6) \times 10^{-5}$& $1.473(6) \times 10^{-5}$ &$-8.7(1.5) \times 10^{-5}$ & $9.5(1.5) \times 10^{-5}$ \\ \hline
        old 59 &$1.471(10) \times 10^{-5}$& $1.471(10) \times 10^{-5}$ &$-6.6(2.5) \times 10^{-5}$ & $7.9(2.3) \times 10^{-5}$ \\ \hline
    \end{tabular}
    \caption{The K to $\pi\pi$ matrix element for Isospin 0 and 2. The I=2 matrix element for $Q_1$ and $Q_2$ are the same because they come from the same three point diagrams.}
    \label{tab:ktopipi}
\end{table}

\subsection{Four-point integrated correlators and $\Delta m_K$}
\begin{figure}

    \centering
        \includegraphics[width=0.4 \textwidth]{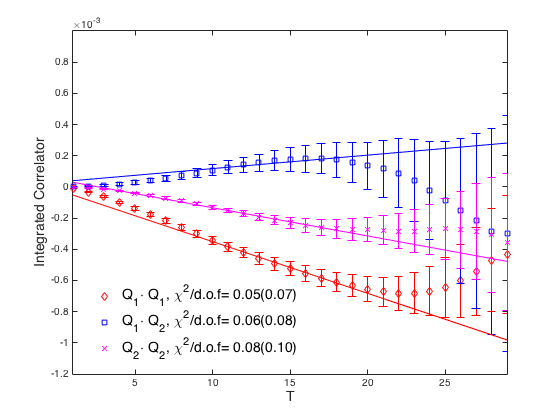}
        \includegraphics[width=0.4 \textwidth]{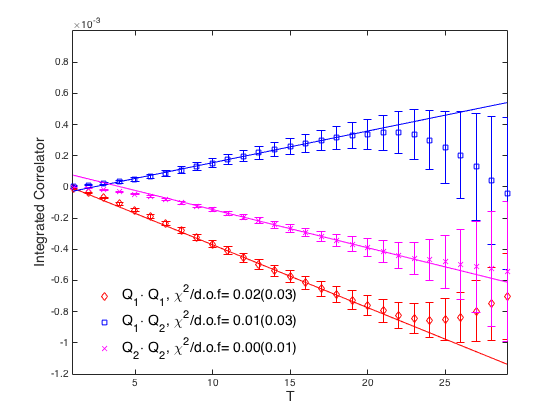}
    \caption{Integrated correlators $\mathcal{A}_{ij}(T)$. The left plot shows the result from "old 59". The right plot shows the result from "new 152".}
    \label{fig:DMK}
\end{figure}

The integrated correlators $\mathcal{A}_{ij}(T)$ are plotted in Figure \ref{fig:DMK} . 
The fitted slopes $k_{ij}$ all should be improved with 2.5 times our earlier statistics. 
The $\Delta m_K$ value and separated contributions from different types of diagrams after normalization are shown in Table \ref{tab:DMK}.
Based on the formula proposed in \cite{finite volume}, our estimated finite-volume correction for $\Delta m_K$ is:
    $$\Delta m_K(FV) = -0.22(7) \times 10^{-12} MeV$$
and our preliminary result for $\Delta m_K$ is:
    $$\Delta m_K  = 7.9(1.3) \times 10^{-12} MeV.$$

In order to realize the GIM mechanism in our calculation, the charm quark is included in our calculation. The lattice spacing in our calculation is $a^{-1} = 2.36 GeV$, which is only twice the charm quark mass. Discretization effects are estimated to be the
largest source of systematic error: 
$\sim (m_c a)^2$ is $ \sim 25 \%$.\\ 

\begin{table}[!htb]
    \centering
    \begin{tabular}{c|c|c|c|c|c } \hline
       Data Set  & $\Delta m_K$ & $\Delta m_K$(tp12) &$\Delta m_K$(tp34)& $\Delta m_K$(tp3) &$\Delta m_K$(tp4)\\ \hline
         new 152 & 8.2(1.3) & 8.3(0.6) & 0.1(1.1) &1.58(31) & -1.28(94) \\ \hline
         old 59 & 5.8(1.8) & 7.0(1.3) & -1.1(1.2) &1.17(43) & -2.16(1.20) \\ \hline
    \end{tabular}
    \caption{Results for $\Delta m_K$ from uncorrelated fits in units of $10^{-12}$ MeV with fitting range 10:20. }
    \label{tab:DMK}
\end{table}

\subsection{Sample AMA correction}


Our use of the sample AMA method reduced the computational cost of the calculation by a factor of 2.3, while the statistical error on the correction will add to the total statistical error. Table \ref{tab:AMA_error} shows the size of the error coming from the correction which is added in quadrature to give our final error. We can conclude that the AMA method does not contribute much to the error in our final answer.

\begin{table}[]
    \centering
    \begin{tabular}{c|c|c|c} \hline
        Data Set &   type 3\&4 error & type 3\&4 error & type 3\&4 error \\ 
                    & from "sloppy" & from correction & in total \\ \hline
        new 152  & 0.9 & 0.6 &  1.1\\ \hline
        old 59 & 1.1 & 0.6  &  1.2 \\ \hline

    \end{tabular}
    \caption{Error contributions to $\Delta m_K$ from type-3 and type-4 diagrams (in units of $10^{-12}$ MeV ). From left to right, type-3 and type-4 errors from "sloppy", from correction and in total are shown. In our calculation, type-3 and type-4 diagrams are AMA corrected while type-1 and type-2 diagrams are only calculated as part of the exact measurements that are also used to determine the corrections for the type-3 and type-4 diagrams.}
    \label{tab:AMA_error}
\end{table}

\section{Conclusion and Outlook}
Our preliminary result for $\Delta m_K$ based on 152 configurations with physical quark masses is:
        $$\Delta m_K  = 7.9(1.3)(2.1) \times 10^{-12} MeV.$$
        
Here the first error is statistical and the second is an estimate of largest systematic error, the discretization error which results from including a charm quark with $m_c = 0.31$ in our calculation. Our $\Delta m_K$ value is to be compared with the experimental value $3.483(6) \times 10^{-12}$ MeV. However, we view such a comparison as premature given the possibly large and poorly estimated finite lattice spacing error. In the future, planned $\Delta m_K$ calculations on SUMMIT with finer lattice spacing will provide a better estimate of the systematic errors coming from discretization effects.

\end{document}